\documentclass[english,jmp,preprint]{revtex4}
\usepackage[T1]{fontenc}
\usepackage[latin1]{inputenc}
\usepackage{amsmath}
\usepackage{amssymb}

\makeatletter
\usepackage{graphicx}

\usepackage{babel}
\makeatother
\begin{document}
\def\figurename{FIG.}

\title{On the Generalized Borel Transform and its Application to the Statistical
Mechanics of Macromolecules. }

\author{Marcelo Marucho}

\email{marucho@polymer.uakron.edu}

\author{Gustavo A. Carri%
\footnote{To whom any correspondence should be addressed%
}}

\email{carri@polymer.uakron.edu}

\affiliation{The Maurice Morton Institute of Polymer Science, The University of
Akron, Akron, OH 44325-3909, USA.}

\begin{abstract}
We present a new integral transform called the Generalized Borel Transform
(GBT) and show how to use it to compute some distribution functions
used to describe the statistico-mechanical behavior of macromolecules.
For this purpose, we choose the Random Flight Model (RFM) of macromolecules
and show that the application of the GBT to this model leads to the
exact expression of the polymer propagator (two-point correlation
function) from which all the statistical properties of the model can
be obtained. We also discuss the mathematical simplicity of the GBT
and its applicability to polymers with other topologies.
\end{abstract}
\maketitle

\section{Introduction}

Consider a group of $n$ small molecules connected in a sequential
manner such that each small molecule has only two nearest neighbours
with whom it forms a chemical bond except for the small molecules
at the end of the chain that have only one connected nearest neighbour.
This chain is called a polymer. In this particular case, the chain
ends do not form a chemical bond thus, the polymer is said to have
a linear topology (linear polymer). If the ends were to form a chemical
bond, then the polymer is said to be a ring (cyclic) polymer. Another
well known topology is the star topology where $m$ linear polymers
grow from the same point and they are connected at this point. This
defines an $m$-arm star polymer. Combinations of these three topologies
or new topologies define more complex macromolecules. 

Another characteristic of polymers is that they can be made of two
or more different kinds of small molecules. 

We conclude this section by speculating about other possible models
for which the GBT might provide some useful insight. But, before we
proceed, we should note that extensions of the present calculation
to other polymer topologies like rings and stars are straightforward.
For example, in the case of rings the cyclic topology of the polymer
generates an extra constraint in the polymer propagator of the form

\begin{equation}
\delta\left(\sum_{j=1}^{n}\mathbf{R}_{j}\right),\label{eq:anillo}\end{equation}
which can be easily handled by the GBT. 

A generalization of Eq. (\ref{eq:1}) that takes into account more
complex systems like copolymers and the effects of external fields
has the mathematical expression given by

\begin{equation}
P\left(\mathbf{R},n,\left\{ p_{j}^{\alpha}\right\} \right)=\int d\left\{ \mathbf{R}_{k}\right\} \prod_{j=1}^{n}\left(\sum_{\alpha=1}^{s}\left(p_{j}^{\alpha}\:\tau_{j}^{\alpha}\left(\mathbf{R}_{j}\right)\right)\right)\delta\left(\sum_{j=1}^{n}\mathbf{R}_{j}-\mathbf{R}\right)\:\exp\left(-\omega\left(\left\{ \mathbf{R}_{j}\right\} \right)\right),\label{eq:copolymers}\end{equation}
where $s$ is the total number of chemical species forming the copolymer,
$p_{j}^{\alpha}$ is the probability of finding the $j$-th segment
in the $\alpha$-th chemical species and $\omega\left(\left\{ \mathbf{R}_{j}\right\} \right)$
has the mathematical form

\begin{equation}
\omega\left(\left\{ \mathbf{R}_{j}\right\} \right)=\sum_{j=1}^{n}\eta\left(\mathbf{R}_{j}\right),\label{eq:potential}\end{equation}
$\eta\left(\mathbf{R}\right)$ being any function. In particular,
the effects of external vectorial $\left(\eta\left(\mathbf{R}\right)=-\mathbf{F}\cdot\mathbf{R}\right)$
and quadrupolar $\left(\eta\left(\mathbf{R}\right)=Q_{ij}R_{i}R_{j}\right)$
fields can be studied exactly. In general, the statistical properties
of the model defined by Eq. (\ref{eq:copolymers}) can be computed
exactly using GBT.

In recent years, new experimental tools have been developed to study
the behavior of single macromolecules. These tools have given researchers
the ability to manipulate polymers individually. Perhaps, the force-elongation
measurements done in different biopolymers using Atomic Force Microscopes
and Optical Tweezers are the best known examples\cite{Molecules}.
This new experimental capability has proved that our most advanced
models are still incomplete and cannot describe modern experimental
data adequately. One example of this statement is the force-elongation
curve of the polyprotein made of the I27 module of human cardiac titin\cite{Molecules_2}.
This curve shows discontinuities that occur when one module of the
polyprotein unravels. The data is generally fitted to the WCM. But,
this model cannot describe structures like $\alpha$-helices or $\beta$-sheets
present in polyproteins. 

The solution of the WCM is mathematically involved and the addition
of secondary structure like $\alpha$-helices does not simplify the
model. On the contrary, the mathematical complexity of this {}``extended''
WCM is expected to be even more involved. As a consequence of this,
we have started to explore new mathematical tools that might lead
to alternative approaches to the models of polymers. One tool that
has proved to be potentially useful is the Generalized Borel Transform
(GBT) which we present in this paper. This technique was borrowed
from Quantum Mechanics (QM) and Quantum Field Theories (QFT) where
it has been successfully applied to solve different problems such
as heavy quark-antiquark interaction in QFT \cite{tecnica2} and singular
scattering potentials in QM \cite{tecnica3}. In addition, this mathematical
tool allows one to obtain analytical solutions of parametric integrals
like Laplace/Mellin transforms for all range of its parameters. Therefore,
it is extremely useful to study non-perturbative regimes. In this
paper, we show how to use this computational tool in the field of
statistical mechanics of single polymers by computing exactly the
polymer propagator of the RFM of linear macromolecules. In the companion
paper, we generalize these ideas to the statistical mechanics of linear
semiflexible polymer chains, as described by the WCM, and present
new results for this model.

The RFM is the simplest attempt to describe the statistical properties
of flexible polymer chains. Thus, it is the first model where the
GBT should be applied. In this model, polymers are assumed to be made
of \textit{n} segments of length \textit{l} (=Kuhn length) which are
connected in a sequential manner. These segments are free to rotate
around any axis as long as the move does not break the polymer chain.
Figure \ref{Figure 1} depicts the RFM.

This paper is organized as follows. In section II we show how is possible
to calculate Fourier sin/cos transforms using the Generalized Borel
Transform and we present a brief summary of the mathematical aspects
of this technique. Then in section III we solve exactly a particular
Fourier sin transform, namely the polymer propagator (normalized distribution
function) of the RFM. Finally, in section IV we present the conclusions
and discuss some other models in polymer physics where this new application
of the GBT might lead to exact solutions.

\section{The Generalized Borel Transform}

In the previous section we have shown that the Fourier Sine transform
plays a very important role in the evaluation of the statistical properties
of models for single macromolecules. Therefore, let us start the description
of the GBT by writing the general expression of a Fourier Sine transform
of a function $H\left(k,a\right)$ 

\begin{equation}
Q\left(R,a\right)=\int_{0}^{\infty}\sin\left(Rk\right)H\left(k,a\right)dk.\label{eq:fouriersin}\end{equation}
Furthermore, consider the Laplace transform of the same function which
is given by the definition

\begin{equation}
S\left(g,a\right)=\int_{0}^{\infty}H\left(x,a\right)\exp\left(-gx\right)dx\qquad g>0.\label{eq:laplace}\end{equation}
Then, we observe that we can obtain the Fourier Sine transform, $Q\left(R,a\right)$,
from the Laplace transfom, $S\left(g,a\right)$, as the analytic continuation
of $S\left(g,a\right)$ to the complex plane as follows

\begin{equation}
Q\left(R,a\right)=Im\left\{ S\left(g=-iR,a\right)\right\} .\label{eq:cont}\end{equation}
Consequently, we will focus on the evaluation of Laplace transforms.
For this purpose, we will employ the GBT technique described hereafter.

The main goal of the GBT is to obtain analytical solutions of parametric
integrals of the Mellin/Laplace type\cite{tecnica1,tecnica2,tecnica3}
for all the range of values of the parameters. Therefore, this technique
is very useful to study non-perturbative regimes. The basic idea of
method consists of introducing two auxiliary functions, $S\left(g,a,n\right)$
and $B_{\lambda}\left(s,a,n\right)$(the Generalized Borel Transform).
These functions depend on auxiliary parameters called $n$ and $\lambda$.
These parameters have no physical meaning and are introduced for the
sole purpose of helping in the computation of an explicit mathematical
expression for $S\left(g,a\right)$ for all values of the true parameters
$g$ and $a$. 

Let us start with the mathematical definition of $S\left(g,a,n\right)$
which is the following

\begin{equation}
S\left(g,a,n\right)=\int_{0}^{\infty}x^{n}H\left(x,a\right)\exp\left(-gx\right)dx.\qquad g>0\label{int}\end{equation}
 We have explicitly extracted a factor $x^{n}$ from the function
to be transformed. This integral is related to the Laplace transform,Eq.
(\ref{eq:laplace}), by the following relationship 

\begin{equation}
S\left(g,a,n\right)=\left(-\right)^{n}\frac{\partial^{n}}{\partial g^{n}}S\left(g,a\right),\label{eq:19}\end{equation}
 which can be inverted to give 

\begin{equation}
S\left(g,a\right)=\left(-\right)^{n}\underbrace{\int dg\cdots\int dg}_{n}S\left(g,a,n\right)+\sum_{p=0}^{n-1}c_{p}\left(a,n\right)g^{p}.\label{fiapro}\end{equation}
 The finite sum comes from the indefinite integrations. Note that
all the coefficients vanish whenever the Laplace transform, Eq. (\ref{eq:laplace}),
fulfills the following asymptotic condition

\begin{equation}
\lim_{g\rightarrow\infty}S\left(g,a\right)=0.\label{eq:asymp}\end{equation}
 In addition, the expression given by Eq. (\ref{fiapro}), is valid
for any value of the parameter $n,$ in particular for $n\gg1$ where
the GBT provides an approximate analytical expression for $S\left(g,a,n\right)$
as we describe below. 

Let us define the Generalized Borel Transform of $S\left(g,a,n\right)$
as follows

\begin{equation}
B_{\lambda}\left(s,a,n\right)\equiv-\int_{0}^{\infty}\exp\left[s/\eta\left(g\right)\right]\left[\frac{1}{\lambda\eta\left(g\right)}+1\right]^{-\lambda s}\frac{S\left(g,a,n\right)}{\left[\eta\left(g\right)\right]^{2}}\frac{\partial\eta\left(g\right)}{\partial g}dg,\quad Re\left(s\right)<0\label{borel}\end{equation}
 where $\lambda$ is any real, positive and non-zero number, and $\eta$
is defined as follows $1/\eta\equiv\lambda\left(\exp\left(g/\lambda\right)-1\right)$.
Then, $B_{\lambda}\left(s,a,n\right)$ is an analytic function for
real values of $s$ less than zero. Moreover, the analytic continuation
to the other half of the complex plane generates an analytic function
with a cut on the positive real axis. 

In order to invert the transform defined by Eq.(\ref{borel}), we
note that the change of variables $u\left(g\right)=1/\eta-\lambda\ln\left[1+1/\lambda\eta\right]$
transforms the integral, Eq.(\ref{borel}), into a Laplace transform

\begin{equation}
B_{\lambda}\left(s,a,n\right)\equiv\int_{0}^{\infty}\exp\left[su\right]L_{\lambda}\left(S,a,n,u\right)du,\quad Re\left(s\right)<0\label{eq:laplaceborel}\end{equation}
where $L_{\lambda}\left(S,a,n,u\right)$ depends on $S\left(g,a,n\right)$.
Consequently, the inverse Laplace transform of Eq.(\ref{eq:laplaceborel})
provides a procedure for the evaluation of $S\left(g,a,n\right)$
by integrating $B_{\lambda}\left(s,a,n\right)$ on the imaginary axis
or over the discontinuity of $B_{\lambda}\left(s,a,n\right)$ on the
cut. After a change of variables we can write $S\left(g,a,n\right)$
as follows

\begin{equation}
S\left(g,a,n\right)=2\lambda^{2}\left(1-\exp\left(-g/\lambda\right)\right)\int_{-\infty}^{\infty}\int_{-\infty}^{\infty}\exp\left[G\left(w,t,g,\lambda,a,n\right)\right]dwdt.\label{doble}\end{equation}
 $G\left(w,t,g,\lambda,a,n\right)$ is given by the following expression
(for more details see Ref. \cite{tecnica2})

\begin{equation}
\begin{array}{c}
{\displaystyle G\left(w,t,g,\lambda,a,n\right)=-s\left(t\right)u\left(g\right)+t-\ln\left\{ \Gamma\left[\lambda\left(s\left(t\right)+x\left(w\right)\right)\right]\right\} }\\
{\displaystyle +\left\{ \lambda\left[s\left(t\right)+x\left(w\right)\right]-1\right\} \ln\left(\lambda s\left(t\right)\right)-\lambda s\left(t\right)+w+\ln\left[x\left(w\right)^{n}H\left(x\left(w\right)\right)\right]},\end{array}\label{eq:G}\end{equation}
where $s\left(t\right)=\lambda\exp\left(t\right)$ and $x\left(w\right)=\exp\left(w\right)$.

Note that the expression given by Eq. (\ref{doble}) is valid for
any non-zero, real and positive value of the parameter $\lambda.$
However, the resulting expression for $S\left(g,a,n\right)$ does
not depend on $\lambda$ explicitly. Thus, each value of the parameter
$\lambda$ defines a particular Borel transform. Consequently, we
can choose the value of this parameter in such a way that it allows
us to solve Eq. (\ref{doble}). 

The dominant contribution to the double integral is obtained using
steepest descent\cite{norman,copson} in the variables $t$ and $w$.
For this purpose, we first compute the expressions of the saddle point
$t_{o}\left(g,a,n\right)$ and $w_{o}\left(g,a,n\right)$ in the limit
$\lambda\gg1.$The results are the following

\begin{equation}
t_{o}\left(g,a,n\right)=\ln\left[\frac{x_{o}^{2}\left(g,a,n\right)}{f\left(x_{o}\left(g,a,n\right),a,n\right)}\right]\quad,\quad w_{o}\left(g,a,n\right)=\ln\left[x_{o}\left(g,a,n\right)\right],\label{tw}\end{equation}
 where $x_{o}\left(g,a,n\right)$ is the real and positive solution
of the implicit equation coming from the extremes of the function
$G\left(w,t,g,\lambda,a,n\right)$ in the asymptotic limit in $\lambda$.
Explicitly, the equation is

\begin{equation}
x_{o}^{2}g^{2}=f\left(x_{o},a,n\right)\left[f\left(x_{o},a,n\right)+1\right],\label{gx}\end{equation}
 where

\begin{equation}
f\left(x_{o},a,n\right)\equiv1+n+x_{o}\frac{d\ln\left[H\left(x_{o},a\right)\right]}{dx_{o}}.\label{fx}\end{equation}

Afterward, we check the positivity condition\cite{jefrey} (the Hessian
of $G\left(w,t,g,\lambda,a,n\right)$ at the saddle point must be
positive). Let us call the Hessian $D\left(x_{o},a,n\right)$. Its
mathematical expression is 

\begin{equation}
D\left(x_{o},a,n\right)\equiv-x_{o}\,\frac{df\left(x_{o},a,n\right)}{dx_{o}}\left[1/2+f\left(x_{o},a,n\right)\right]+f\left(x_{o},a,n\right)\left[1+f\left(x_{o},a,n\right)\right].\label{dx}\end{equation}

Observe that in the range of the parameters where $f\left(x_{o},a,n\right)\gg1$,
which is fulfilled when $n\gg1$, we can keep the second order term
in the expansion of $G\left(w,t,g,\lambda,a,n\right)$ around the
saddle point. Then, we can approximate the double integral in Eq.
(\ref{doble}) as follows

\begin{equation}
S_{Aprox}\left(g,a,n\right)=4\pi\frac{\lambda^{2}\left(1-\exp\left(-g/\lambda\right)\right)}{\sqrt{D\left[x_{o},a,n\right]}}\exp\left[G\left(w_{o},t_{o},g,\lambda,a,n\right)\right].\label{eq:infini}\end{equation}

In the limit $\lambda\rightarrow\infty$ we obtain the following approximate
expression for $S\left(g,a,n\right)$

\begin{equation}
S_{Aprox}\left(g,a,n\right)=\sqrt{2\pi}e^{-1/2}\frac{\sqrt{f\left[x_{o},a,n\right]+1}}{\sqrt{D\left[x_{o},a,n\right]}}\left[x_{o}\right]^{n+1}H\left[x_{o},a\right]\exp\left[-f\left[x_{o},a,n\right]\right].\label{eq:res}\end{equation}

Note that the expression given by Eq. (\ref{eq:res}) is valid for
functions $H\left(x,a\right)$ that fulfill the following general
conditions. Firstly, the relationship given by Eq. (\ref{gx}) must
be biunivocal. Secondly, $D\left(x_{o},a,n\right)$ must be positive
at $x_{o}$. Thirdly, $f\left(x_{o},a,n\right)$ must be larger than
one. In particular, this condition is fulfilled when $n\gg1.$ These
conditions provide the range of values of the parameters where the
approximate solution, Eq. (\ref{eq:res}), is valid. 

Finally, we replace the expression given by Eq. (\ref{eq:res}) into
Eq. (\ref{fiapro}) to obtain an approximate analytical expression
for the Laplace transform $S\left(g,a\right)$. In particular, in
the limit $n\rightarrow\infty$, we obtain the following analytical
solution for $S\left(g,a\right)$ 

\begin{equation}
S\left(g,a\right)=\begin{array}{c}
\lim\\
n\rightarrow\infty\end{array}\left(-\right)^{n}\underbrace{\int dg\cdots\int dg}_{n}S_{Aprox}\left(g,a,n\right).\label{fff}\end{equation}

One particular case of this result is the one where $H\left(x,a\right)$
does not contribute to the saddle point. This is the case when $f\left(x_{o},a,n\right)$
can be approximated by $1+n$ (the derivative of $H\left(x_{o},a\right)$
is negligible). Then, the saddle point solution is $x_{o}\left(g,a,n\right)\simeq\left(n+3/2\right)/g$
and the expression of $S_{Aprox}\left(g,a,n\right)$ is

\begin{equation}
S_{Aprox}\left(g,a,n\right)\simeq\frac{\Gamma\left(n+1\right)}{g^{n+1}}H\left[x_{o}\left(g,a,n\right),a\right]\quad n\gg1.\label{eq:Sapo}\end{equation}

Another important property of the expression given by Eq. (\ref{fff})
is that, in the limit $n\rightarrow\infty$, the approximate solution,
Eq. (\ref{eq:res}), becomes an exact solution for Eq. (\ref{int}).
Thus, as long as the $n$ indefinite integrals are calculated without
approximations, then Eq. (\ref{fff}) is an exact solution for Eq.
(\ref{eq:laplace}). 

In summary, the procedure to use the GBT to compute Fourier Sine transforms
is the following. Firstly, one has to solve the implicit equation,
Eq. (\ref{gx}), for $n\gg1$ to obtain the mathematical expression
of $x_{o}\left(g,a,n\right)$. Replacing this expression into Eq.
(\ref{eq:res}) and doing the $n$ indefinite integrals in Eq. (\ref{fff}),
we get the exact/approximate expression for $S\left(g,a\right)$.
Finally, one has to compute the analytic continuation of $S\left(g,a\right)$,
Eq. (\ref{eq:cont}), to get the exact/approximate solution of the
Fourier Sine transform, Eq. (\ref{eq:fouriersin}). 

In the next section we apply this technique to solve exactly the statistical
mechanics of flexible macromolecules.

\section{Application}

Let us start by analyzing the polymer propagator predicted by the
Random-Flight Model which is given by

\begin{equation}
P\left(\mathbf{R},n\right)=\int d\left\{ \mathbf{R}_{k}\right\} \prod_{j=1}^{n}\tau\left(\mathbf{R}_{j}\right)\delta\left(\sum_{j=1}^{n}\mathbf{R}_{j}-\mathbf{R}\right),\label{eq:1}\end{equation}
 where $\mathbf{R}_{j}$ is the bond vector between the $(j-1)$-th
and $j$-th beads, $n$ is the total number of segments per polymer
chain, $\mathbf{R}$ is the end-to-end vector and $\tau\left(\mathbf{R}_{j}\right)$
is given by the formula

\begin{equation}
\tau\left(\mathbf{R}_{j}\right)=\frac{\delta\left(\left|\mathbf{R}_{j}\right|-l\right)}{4\pi l^{2}}.\label{eq:2}\end{equation}

Using the Fourier representation of the delta function, we obtain

\begin{equation}
\begin{array}{c}
{\displaystyle P\left(\mathbf{R},n\right)=\int\frac{d^{3}k\exp\left(-i\mathbf{R}\cdot\mathbf{k}\right)}{\left(2\pi\right)^{3}\left(4\pi l^{2}\right)^{n}}\left[\int d\left\{ \mathbf{R}_{k}\right\} \prod_{j=1}^{n}\delta\left(\left|\mathbf{R}_{j}\right|-l\right)\exp\left(i\sum_{j=1}^{n}\mathbf{R}_{j}\cdot\mathbf{k}\right)\right]}\\
{\displaystyle =\int\frac{d^{3}k}{\left(2\pi\right)^{3}}\exp\left(-i\mathbf{R}\cdot\mathbf{k}\right)K\left(\mathbf{k},n,l\right),}\end{array}\label{eq:distri}\end{equation}
 where the characteristic function, $K\left(\mathbf{k},n,l\right)$,
is

\begin{equation}
K\left(\mathbf{k},n,l\right)=\left(\frac{\sin\left(\left|\mathbf{k}\right|l\right)}{\left|\mathbf{k}\right|l}\right)^{n}.\label{eq:K}\end{equation}

The evaluation of the angular integrals in Eq. (\ref{eq:distri})
is straightforward. After rescaling $\mathbf{R}$ and $\mathbf{k}$
with the Kuhn length, $l$, we obtain the final expression for the
polymer propagator

\begin{equation}
P\left(R,n\right)=\frac{2}{\left(2\pi\right)^{2}R}\int_{0}^{\infty}dk\left[\sin\left(kR\right)\left(\frac{\sin\left(k\right)}{k}\right)^{n}k\right],\label{eq:dis}\end{equation}
where $k=\left|\mathbf{k}\right|$ and $R=\left|\mathbf{R}\right|$.

This integral representation of the polymer propagator is a Fourier
Sine transform and can be solve exactly using GBT. Then, our first
step consists of expressing the polymer propagator, Eq. (\ref{eq:dis}),
in terms of a Laplace Transform. For this purpose we define the function

\begin{equation}
S\left(b,n\right)\equiv\int_{0}^{\infty}dw\left[\exp\left(-wb\right)\left(\frac{\sin\left(w\right)}{w}\right)^{n}w\right].\label{eq:gb}\end{equation}
 from where we recover the expression of the polymer propagator, Eq.
(\ref{eq:dis}), as the analytic continuation of the function $S\left(b,n\right)$
to the complex plane

\begin{equation}
P\left(R,n\right)=\frac{2}{\left(2\pi\right)^{2}R}Im\left\{ S\left(b=-iR,n\right)\right\} .\label{eq:disg}\end{equation}

Let us now rewrite Eq. (\ref{eq:gb}) as follows

\begin{equation}
S\left(b,n\right)=\frac{\partial^{n}}{\partial c^{n}}\left\{ \int_{0}^{\infty}dw\left[w\exp\left(-wb\right)\exp\left(c\frac{\sin\left(w\right)}{w}\right)\right]\right\} _{c=0}=\frac{\partial^{n}}{\partial c^{n}}\left\{ GA\left(b,c\right)\right\} _{c=0},\label{eq:gb2}\end{equation}
 where

\begin{equation}
GA\left(b,c\right)\equiv\int_{0}^{\infty}dww\exp\left(-wb\right)H\left(w,c\right),\label{eq:ga}\end{equation}
 and

\begin{equation}
H\left(w,c\right)\equiv\exp\left[c\frac{\sin\left(w\right)}{w}\right].\label{eq:3}\end{equation}
 Then, the integral expressed by Eq. (\ref{eq:ga}) satisfies all
the requirements of the GBT technique \cite{tecnica3}. Consequently,
we evaluate this integral in the following way

\begin{equation}
GA\left(b,c\right)=\begin{array}{c}
\lim\\
N\rightarrow\infty\end{array}\begin{array}{c}
\left(-\right)^{N}\underbrace{\int db\cdots\cdots\cdots\int db}GA_{N},\\
N\end{array}\label{eq:gaprox}\end{equation}
 where we have defined

\begin{equation}
GA_{N}\left(b,c\right)\equiv\int_{0}^{\infty}dw\left[w^{N+1}\exp\left(-wb\right)H\left(w,c\right)\right].\label{eq:gaene}\end{equation}

In the asymptotic limit of $N\rightarrow\infty$, the GBT provides
an analytical solution for Eq. (\ref{eq:gaene}). Following the technique,
we first solve the following implicit equation for $w_{o}$, Eq. (\ref{gx}),

\begin{equation}
\left\{ N+1+w_{o}\frac{\partial}{\partial w_{o}}\ln\left[H\left(w_{o},c\right)\right]\right\} \left\{ N+2+w_{o}\frac{\partial}{\partial w_{o}}\ln\left[H\left(w_{o},c\right)\right]\right\} =w_{o}^{2}b^{2},\label{eq:4}\end{equation}
 whose asymptotic solution is

\begin{equation}
w_{o}\simeq\frac{N+5/2}{b}\quad N\gg1.\label{eq:sp}\end{equation}
 Replacing this expression for $w_{o}$ in the expression provided
by the GBT, Eq. (\ref{eq:res}), we obtain

\begin{equation}
GA_{N}\left(b,c\right)\simeq\frac{\Gamma\left(N+2\right)}{b^{N+2}}H\left(\frac{N+5/2}{b},c\right)\quad N\gg1.\label{eq:gan}\end{equation}
 Furthermore, we replace Eq. (\ref{eq:gan}) into Eq. (\ref{eq:gaprox})
and the resulting expression into Eq. (\ref{eq:gb2}), then we obtain

\begin{align}
S\left(b,n\right) & =\begin{array}{c}
\lim\\
c\rightarrow0\end{array}\frac{\partial^{n}}{\partial c^{n}}\left\{ \begin{array}{c}
\lim\\
N\rightarrow\infty\end{array}\begin{array}{l}
\left(-\right)^{N}\underbrace{\int db\cdots\cdots\cdots\int db}\frac{\Gamma\left(N+2\right)}{b^{N+2}}H\left(\frac{N+5/2}{b},c\right)\\
\qquad\quad\qquad\quad N\end{array}\right\} .\label{eq:5}\end{align}

We now proceed to exchange the order of the operators, first we evaluate
the $n$-th derivative of the function $H$ with respect to $c$ and,
afterward, we take the limit of $c\rightarrow0$ . As a result, we
obtain

\begin{equation}
S\left(b,n\right)=\begin{array}{c}
\lim\\
N\rightarrow\infty\end{array}\left(-\right)^{N}\int db\cdots\int db\frac{\Gamma\left(N+2\right)}{b^{N+2}}\left(\frac{\sin\left(\frac{N}{b}\right)}{\left(\frac{N}{b}\right)}\right)^{n}.\label{eq:6}\end{equation}
 Next, we solve the $N$ integrations. Using properties of the function
$\sin\left(x\right)$ we can write $S\left(b,n\right)$ for any odd
number of segments as follows

\begin{equation}
S\left(b,n\right)=\frac{1}{2^{n-1}}\sum_{k=0}^{\frac{n-1}{2}}\left(-\right)^{\frac{n-1}{2}+k}\left(\begin{array}{c}
n\\
k\end{array}\right)M\left(N,n,k,b\right),\label{eq:gb3}\end{equation}
 where

\begin{equation}
M\left(N,n,k,b\right)\equiv\begin{array}{c}
\lim\\
N\rightarrow\infty\end{array}\left(-\right)^{N}Im\sum_{r=0}^{\infty}\frac{\left(i\left(n-2k\right)\right)^{r}N^{r-n}}{r!}\int db\cdots\int db\frac{\Gamma\left(N+2\right)}{b^{N+2-n+r}}.\label{eq:eme}\end{equation}

We note that the only powers of $b$ in Eq. (\ref{eq:eme}) that fulfill
the asymptotic behavior of the function $S\left(b,n\right)$, Eq.
(\ref{eq:asymp}), are those that satisfy the condition $r\geq\left(n-1\right)$.
Consequently, the $N$ indefinite integrations are exactly doable,
the result is

\begin{equation}
\int db\cdots\int db\frac{1}{b^{N+2-n+r}}=\frac{\Gamma\left(2+r-n\right)}{\Gamma\left(N+2-n+r\right)}\frac{\left(-\right)^{N}}{b^{2+r-n}}.\label{eq:nint}\end{equation}
 Replacing Eq. (\ref{eq:nint}) into Eq. (\ref{eq:eme}) and introducing
the dummy variable $r=x+n-1$ we can write

\begin{equation}
\begin{array}{c}
{\displaystyle M\left(N,n,k,b\right)\equiv Im\frac{1}{b}\left(i\left(n-2k\right)\right)^{n-1}\sum_{x=0}^{\infty}\left(\frac{i\left(n-2k\right)}{b}\right)^{x}\frac{\Gamma\left(x+1\right)}{\Gamma\left(x+n\right)}}\\
{\displaystyle \times\begin{array}{c}
\lim\\
N\rightarrow\infty\end{array}\frac{N^{x-1}\Gamma\left(N+2\right)}{\Gamma\left(N+x+1\right)},}\end{array}\label{eq:eme2}\end{equation}
 which, after using the asymptotic properties of the Gamma function
\cite{abramovich}

\begin{equation}
\begin{array}{c}
\lim\\
N\rightarrow\infty\end{array}\frac{N^{x-1}\Gamma\left(N+2\right)}{\Gamma\left(N+x+1\right)}=1,\label{eq:7}\end{equation}
 becomes

\begin{equation}
M\left(n,k,b\right)=\frac{1}{b}Im\sum_{x=0}^{\infty}\left(i\left(n-2k\right)\right)^{n-1}\left(\frac{i\left(n-2k\right)}{b}\right)^{x}\,\frac{\Gamma\left(x+1\right)}{\Gamma\left(x+n\right)}.\label{eq:8}\end{equation}

The sum over $x$ is doable, the result gives the following expression
for $M\left(n,k,b\right)$

\begin{equation}
M\left(n,k,b\right)=\frac{1}{b}Im\left[\left(i\left(n-2k\right)\right)^{n-1}FD\left(n,k,b\right)\right],\label{eq:eme3}\end{equation}
 where we have defined

\begin{equation}
\begin{array}{c}
{\displaystyle FD\left(n,k,b\right)\equiv\frac{\Gamma\left(\frac{1}{2}\right)\,_{3}F_{2}\left(\left[1,1,\frac{1}{2}\right],\left[\frac{n+1}{2},\frac{n}{2}\right],-\frac{\left(n-2k\right)^{2}}{b^{2}}\right)}{\sqrt{\pi}\Gamma\left(n\right)}}\\
{\displaystyle +\frac{i\left(n-2k\right)}{b}\frac{\,_{3}F_{2}\left(\left[1,1,\frac{3}{2}\right],\left[\frac{n+1}{2},\frac{n+2}{2}\right],-\frac{\left(n-2k\right)^{2}}{b^{2}}\right)}{\Gamma\left(n+1\right)}.}\end{array}\label{eq:9}\end{equation}
 $_{3}F_{2}\left(\left[,,\right],\left[,\right],x\right)$ is the
Generalized Hypergeometric function \cite{hyper}. From Eq. (\ref{eq:eme3})
we can see that the imaginary part affects only the function $FD\left(n,k,b\right)$.
Thus, we obtain the final expression for $S\left(b,n\right)$

\begin{equation}
S\left(b,n\right)=\sum_{k=0}^{\frac{n-1}{2}}\left(-\right)^{k}\left(\begin{array}{c}
n\\
k\end{array}\right)\frac{\left(n-2k\right)^{n}{}_{3}F_{2}\left(\left[1,1,\frac{3}{2}\right],\left[\frac{n+1}{2},\frac{n+2}{2}\right],-\frac{\left(n-2k\right)^{2}}{b^{2}}\right)}{b^{2}2^{n-1}\Gamma\left(n+1\right)}.\label{eq:gbfinal}\end{equation}

The last step to obtain the analytical expression of the polymer propagator
consists of inserting Eq. (\ref{eq:gbfinal}) into Eq. (\ref{eq:disg})
and computing the analytic continuation of the resulting expression
to the complex plane through the substitution $b=-iR$. After doing
these computations, we arrived at the following expression for the
polymer propagator

\begin{equation}
\begin{array}{c}
{\displaystyle P\left(R,n\right)=\frac{1}{2^{n}\pi^{2}R^{3}}\sum_{k=0}^{\frac{n-1}{2}}\left(-\right)^{k+1}\left(\begin{array}{c}
n\\
k\end{array}\right)\left(n-2k\right)^{n}\frac{1}{\Gamma\left(n+1\right)}}\\
{\displaystyle \times Im\left\{ _{3}F_{2}\left(\left[1,1,\frac{3}{2}\right],\left[\frac{n+1}{2},\frac{n+2}{2}\right],\frac{\left(n-2k\right)^{2}}{R^{2}}\right)\right\} .}\end{array}\label{eq:disim}\end{equation}

This expression can be simplified even further if we use the well
known analytical properties of the Hypergeometric function $_{3}F_{2}\left(z\right)$
\cite{hyper} which is an analytic function for values of $\left|z\right|<1$
and its continuation to the rest of complex plane generates one cut
on the positive real axis starting at $Re\left(z\right)=1$. This
implies that only values of $\frac{\left(n-2k\right)^{2}}{R^{2}}\geq1$
will contribute to the imaginary part of $_{3}F_{2}\left(z\right)$.
Consequently, this condition reduces the number of terms in the $k$-sum
such that the last term of Eq. (\ref{eq:disim}) is $k=\left[\frac{n-R}{2}\right]$.

The explicit evaluation of $Im\left\{ _{3}F_{2}\left(\left[1,1,\frac{3}{2}\right],\left[\frac{n+1}{2},\frac{n+2}{2}\right],\frac{\left(n-2k\right)^{2}}{R^{2}}\right)\right\} $
can be found in Appendix A. The final expression is

\begin{equation}
\begin{array}{c}
{\displaystyle Im\left\{ _{3}F_{2}\left(\left[1,1,\frac{3}{2}\right],\left[\frac{n+1}{2},\frac{n+2}{2}\right],\frac{\left(n-2k\right)^{2}}{R^{2}}\right)\right\} }\\
{\displaystyle =-\frac{R^{2}\pi}{2\left(n-2k\right)^{n}}\frac{\Gamma\left(n+1\right)}{\Gamma\left(n-1\right)}\left[n-2k-R\right]^{n-2}.}\end{array}\qquad n\geq2\label{eq:hyperim}\end{equation}

Finally, we replace Eq. (\ref{eq:hyperim}) into Eq.(\ref{eq:disim})
to obtain the exact expression for the polymer propagator

\begin{equation}
P\left(R,n\right)=\frac{1}{2^{n+1}\pi R}\sum_{k=0}^{\left[\frac{n-R}{2}\right]}\left(-\right)^{k}\left(\begin{array}{c}
n\\
k\end{array}\right)\frac{\left[n-2k-R\right]^{n-2}}{\Gamma\left(n-1\right)}.\label{eq:disfin}\end{equation}

Equation (\ref{eq:disfin}) is valid for odd number of segments but,
it is extended to polymers with any number of segments larger than
two via analytic continuation. 

Therefore, we have obtained the well-known \cite{Wang,Hs} exact analytical
expression for the polymer propagator of flexible chains, Eq. (\ref{eq:dis}),
with any number of segments, $n$, and any end-to-end distance, $R$. 

Observe that the result given by Eq. (\ref{eq:disfin}) can be used
to describe the statistical properties of polymers with other topologies.
For example, consider the case of a flexible $m$-arm star polymer
as shown in Fig. . Since the polymer is flexible, then each arm behaves
independently from the other ones except for the fact that all of
them start at the origin. Thus, the probability of finding the end
of the $j$-th arm in the shell of radius $R_{j}$ with thickness
$dR_{j}$ centered at the origin is 

\begin{equation}
4\pi R_{j}^{2}P\left(R_{j},n_{j}\right)dR_{j},\label{eq:star}\end{equation}
where $n_{j}$ is the number of segments in the $j$-th arm. If we
consider all the arms, then the probability of finding the end of
the first, second, etc. arms in the shells of radii $R_{1}$,$R_{2}$,
etc. with thicknesses $dR_{1}$, $dR_{2}$, etc. centered at the origin
is

\begin{equation}
\left(4\pi\right)^{m}\prod_{j=1}^{m}R_{j}^{2}P\left(R_{j},n_{j}\right)dR_{j}.\label{eq:manystar}\end{equation}
Other probability distributions for star polymers can also be computed
easily. 

Another example is the case of ring (cyclic) polymers. Figure shows
this topology. From this figure and following the steps presented
in this paper for linear polymers it can be proved that the probability
of finding any pair of segments separated by a distance $R$ should
be proportional to the product of two propagators of the form given
by Eq. (\ref{eq:disfin})

\begin{equation}
P_{Ring}\left(R,s,L-s\right)\propto P\left(R,s\right)\: P\left(R,n-s\right),\label{eq:ring}\end{equation}
where $n$ is the total number of segments in the ring and $s$ is
the number of segments (along the contour of the polymer chain) between
the two chosen segments.

The aformentioned two examples clearly show that the results obtained
for linear polymers using the GBT can be used for polymers with other
topologies thus, increasing the range of models that are mathematically
tractable with the GBT.

\section{conclusions}

In this paper we have described a new mathematical method called the
Generalized Borel Transform and applied it to compute some statistical
properties (polymer propagator) of models of flexible polymers. Specifically,
we showed how the GBT is constructed and how to use it to compute
Mellin/Laplace transforms. Moreover, some mathematical properties
were presented. The application of this technique to the statistical
mechanics of single flexible polymers led to the exact solution for
the polymer propagator of linear polymers. The propagator obtained
turned out to be a finite sum of polynomials valid for any end-to-end
distance, $R$, and number of segments, $n$. Furthermore, this result
was used to construct distribution functions for two other topologies,
rings and stars. 

The exact computation of the polymer propagator of the RFM is a straightforward
calculation that requires simple mathematics when the GBT is used.
This mathematical simplicity of the GBT makes it a potentially very
useful computational tool for more complex models of single polymer
chains because it does not add any complexity to the physics of the
starting model. 

Equation (\ref{eq:disfin}) together with its extensions to stars
and rings, Eqs. (\ref{eq:manystar}) and (\ref{eq:ring}), and the
discussion presented in the introduction show that the GBT can solve
exactly a wide range of models for polymers. However, more advanced
models of single polymer chains like the Wormlike Chain Model or helical
polymers where the bond vectors are correlated with each other through
potential interactions are not exactly soluble with the GBT at present.
A generalization of the GBT to multidimensional integrals is required
to address these models.

\section*{ACKNOWLEDGMENTS}

We acknowledge the National Science Foundation, Grant \# CHE-0132278
(CAREER), the Ohio Board of Regents Action Fund, Proposal \# R566
and The University of Akron for financial support.

\appendix

\section{evaluation of the Imaginary part of the Hypergeometric function}

In this appendix, we calculate the expression $Im\left\{ _{3}F_{2}\left(z\right)\right\} $.
For this purpose, we use the following integral representation of
the Hypergeometric function \cite{tablarusa}

\begin{equation}
\begin{array}{c}
{\displaystyle _{3}F_{2}\left(\left[-\nu,\frac{\lambda}{2},\frac{\lambda+1}{2}\right],\left[\frac{\lambda+\mu}{2},\frac{\lambda+\mu+1}{2}\right],-\frac{q^{2}}{w^{2}}\right)=\left[w^{2\gamma}q^{\lambda+\mu-1}B\left(\lambda,\mu\right)\right]^{-1}}\\
{\displaystyle {\times\int_{0}^{q}x^{\lambda-1}\left[q-x\right]^{\mu-1}\left[x^{2}+w^{2}\right]^{\nu}dx,}\quad\lambda,\mu>0,\; Re\left(\frac{q}{w}\right)>0}\end{array}\label{eq:20}\end{equation}
where $B\left(\lambda,\mu\right)$ is the Beta function \cite{tablarusa}.

We now assign the values $\nu=-1,\;\lambda=2,\;\mu=n-1,\; q=n-2k,\; w=b$
to the parameters in Eq. (\ref{eq:20}) to obtain

\begin{equation}
\begin{array}{c}
{\displaystyle _{3}F_{2}\left(\left[1,1,\frac{3}{2}\right],\left[\frac{n+1}{2},\frac{n+2}{2}\right],-\frac{\left(n-2k\right)^{2}}{b^{2}}\right)=\frac{1}{b^{-2}\left(n-2k\right)^{n}B\left(2,n-1\right)}}\\
{\displaystyle \times\int_{0}^{n-2k}x\left[n-2k-x\right]^{n-2}\left[x^{2}+b^{2}\right]^{-1}dx.}\end{array}\label{eq:21}\end{equation}
This integral representation is valid only for $n\geq2$. Therefore,
when we take the analytic continuation to the complex plane as before
$(b=-iR)$, we can express the imaginary part of the Hypergeometric
function as follows

\begin{equation}
\begin{array}{c}
{\displaystyle Im\left\{ _{3}F_{2}\left(\left[1,1,\frac{3}{2}\right],\left[\frac{n+1}{2},\frac{n+2}{2}\right],\frac{\left(n-2k\right)^{2}}{R^{2}}\right)\right\} =-\frac{R^{2}}{\left(n-2k\right)^{n}B\left(2,n-1\right)}}\\
{\displaystyle \times Im\int_{0}^{n-2k}x\left[n-2k-x\right]^{n-2}\left[x^{2}-R^{2}\right]^{-1}dx.}\end{array}\label{eq:imaginariahyper}\end{equation}
Thus, we need to evaluate

\begin{equation}
L\equiv Im\int_{0}^{n-2k}x\left[n-2k-x\right]^{n-2}\left[x-R\right]^{-1}\left[x+R\right]^{-1}dx.\label{eq:22}\end{equation}
 After analyzing the analytical behavior of the integrand, we concluded
that we can exchange the operations of integration and imaginary part
to obtain

\begin{equation}
L=\int_{0}^{n-2k}x\left[n-2k-x\right]^{n-2}\left[x+R\right]^{-1}Im\left\{ \left[x-R\right]^{-1}\right\} dx.\label{eq:ele}\end{equation}
 Thus, we have to compute

\begin{equation}
LS=Im\left\{ \frac{1}{\left(x-R\right)}\right\} ,\label{eq:23}\end{equation}
 first and, afterward, we have to solve the integral given by Eq.
(\ref{eq:ele}).

The analytical behavior of the function $\left(x-R\right)^{-1}$ is
well known. It is an analytic function for $\left|x\right|>R$ but,
its analytic continuation to the complex plane generates a cut on
the real axis in the range $-R<Re\left(x\right)<R$. This cut generates
its imaginary part which is

\begin{equation}
Im\left\{ \frac{1}{\left(x-R\right)}\right\} =\pi\delta\left(x-R\right).\label{eq:delta2}\end{equation}
 Thus, replacing the expression given by Eq. (\ref{eq:delta2}) into
Eq. (\ref{eq:ele}) and making the change of variables $y=x-R$, we
obtain

\begin{equation}
L=\pi\int_{-R}^{n-2k-R}F_{k}\left(y,n,R\right)\delta\left(y\right)dy,\label{eq:24}\end{equation}
 where

\begin{equation}
F_{k}\left(y,n,R\right)\equiv\left(y+R\right)\left[n-2k-R-y\right]^{n-2}\left[y+2R\right]^{-1}.\label{eq:25}\end{equation}
 The result of the integration gives

\begin{equation}
L=\frac{\pi}{2}\left[n-2k-R\right]^{n-2}.\label{eq:elegrande}\end{equation}

Finally, we replace Eq. (\ref{eq:elegrande}) into Eq. (\ref{eq:imaginariahyper})
to obtain the final expression

\begin{equation}
\begin{array}{c}
{\displaystyle Im\left\{ _{3}F_{2}\left(\left[1,1,\frac{3}{2}\right],\left[\frac{n+1}{2},\frac{n+2}{2}\right],\frac{\left(n-2k\right)^{2}}{R^{2}}\right)\right\} }\\
{\displaystyle =-\frac{\pi}{2}\frac{R^{2}}{\left(n-2k\right)^{n}B\left(2,n-1\right)}\left[n-2k-R\right]^{n-2}.}\end{array}\label{eq:26}\end{equation}

\clearpage\subsection*{List of Figures}

\begin{figure}[htbp]
\includegraphics[scale=0.6]{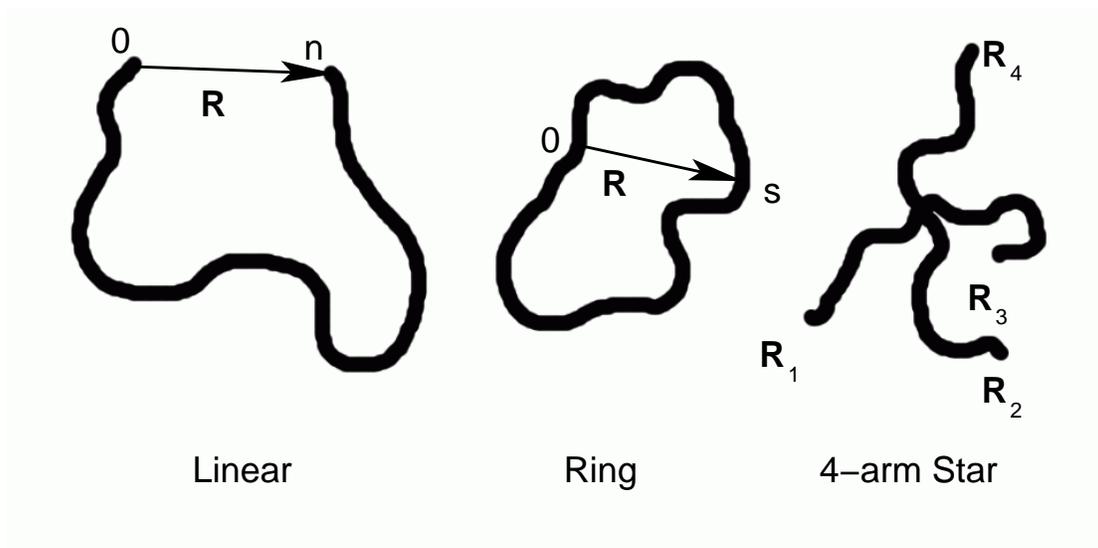}

\caption{Linear, ring and 4-arm star topologies for flexible polymers. $\mathbf{R}$
indicates the end-to-end vector for the linear topology and the relative
position of two segments in the case of a ring polymer. $\mathbf{R}_{j}$
indicates the position of the end of the $j$-th arm in the star topology.}

\label{Figure 1}
\end{figure}

\begin{figure}[htbp]
\includegraphics[scale=0.6]{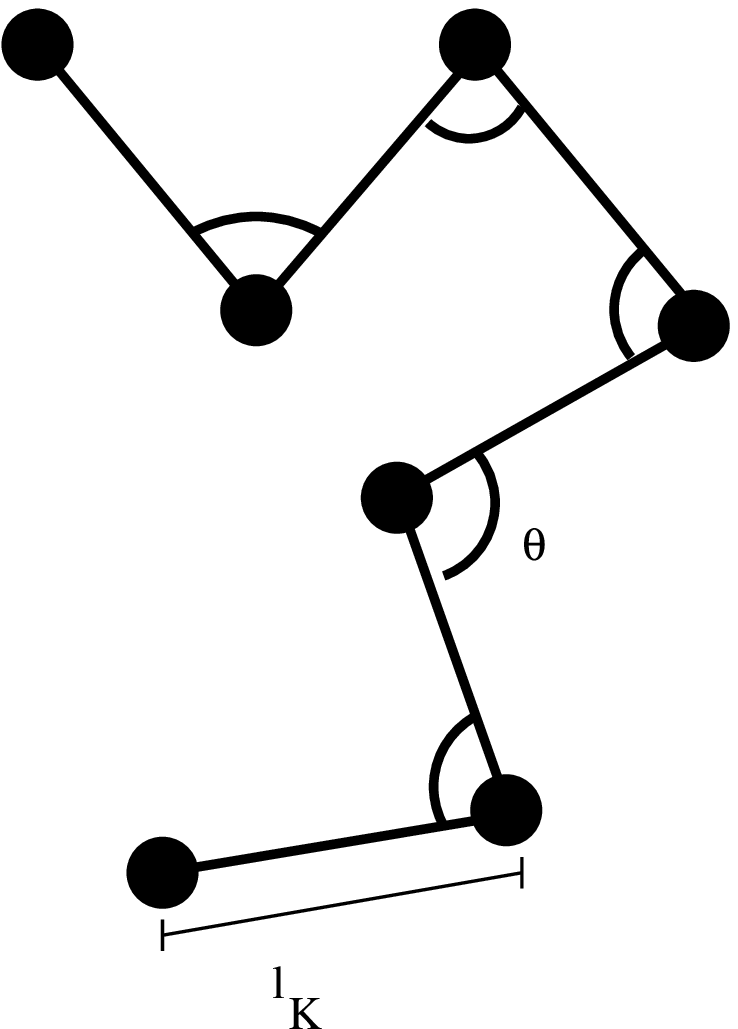}

\caption{The Random Flight Model of polymer chains. $l$ is the Kuhn length
and $\theta$ is the bond angle.}

\label{Figure 2}
\end{figure}

\begin{itemize}

\item[FIG. \ref{Figure 1}:] Linear, ring and 4-arm star topologies for flexible polymers. $\bf{R}$ indicates the end-to-end vector for the linear topology and the relative position of two segments in the case of a ring polymer. $\bf{R}_j$ indicates the position of the end of the $j$-th arm in the star topology. 

\item[FIG. \ref{Figure 2}:] The Random Flight Model of polymer chains. $l$ is the Kuhn length and $\theta$ is the bond angle.

\end{itemize}

\clearpage

\clearpage

\clearpage

\clearpage

\begin{thebibliography}{10}
\bibitem{Kuhn}W. Kuhn, \textit{Kolloid Z}. \textbf{68}, 2 (1934).
\bibitem{desCloiseaux}J. des Cloiseaux and G. Jannink, \textit{Polymers in Solution, Their
Modeling and Structure} (Clarendon Press, Oxford, 1990).
\bibitem{Freed}K. F. Freed, \textit{Renormalization Group Theory of Macromolecules}
(John Wiley \& Sons, New York, 1987). 
\bibitem{Yamakawa_1}H. Yamakawa, \textit{Modern Theory of Polymer Solutions} (Harper \&
Row, New York, 1971). 
\bibitem{Yamakawa_2}H. Yamakawa, \textit{Helical Worm-like Chains in Polymer Solutions}
(Springer, Berlin, 1997). 
\bibitem{Mattice}W. L. Mattice and U. W. Suter, \textit{Conformational Theory of Large
Macromolecules: The Rotational Isomeric State Model in Macromolecular
Systems} (John Wiley \& Sons, Ner York, 1994). 
\bibitem{Flory}P. J. Flory, \textit{Statistical Mechanics of Chain Molecules} (Oxford
University Press, New York, 1988). 
\bibitem{Fujita}H. Fujita, \textit{Polymer Solutions} (Elsevier, New York, 1990). 
\bibitem{Muthu_2}M. Muthukumar and S. F. Edwards, \textit{J. Chem. Phys.} \textbf{76},
2720 (1982). 
\bibitem{Freed_2}W. E. McMullen and K. F. Freed, \textit{Macromolecules} \textbf{23},
255 (1990); W. E. McMullen and K. F. Freed, \textit{J. Chem. Phys.}
\textbf{92},1413 (1990); H. Tang and K. F. Freed, \textit{J. Chem.
Phys.} \textbf{94}, 1572 (1991). 
\bibitem{Copolymers}L. Leiber, \textit{Macromolecules} \textbf{13}, 1602 (1980); T. Otha
and K. Kawasaki, \textit{Macromolecules} \textbf{19}, 2621 (1986);
G. H. Fredrickson and E. Helfand, \textit{J. Chem. Phys.} \textbf{87},
697 (1987); J. L. Barrat and G. H. Fredrickson, \textit{J. Chem. Phys.}
\textbf{95}, 1281 (1991). 
\bibitem{Mixture}L. Kielhorn and M. Muthukumar, \textit{J. Chem. Phys.} \textbf{107},
5588 (1997); D. Broseta and G. H. Fredrickson, \textit{J. Chem. Phys.}
\textbf{93}, 2927 (1990). 
\bibitem{ed}S. F. Edwards, \textit{Proc. Phys. Soc.} \textbf{85}, 613 (1965),
S. F. Edwards, \textit{ibid}. \textbf{88}, 265 (1966).
\bibitem{Molecules}M. G. Poirier \textit{et al.}, \textit{Phys. Rev. Lett.} \textbf{86},
360 (2001); J. F. Leger \textit{et at.}, \textit{Phys. Rev. Lett.}
\textbf{83}, 1066 (1999); V. Parpura and J. M. Fernandez, \textit{Biophys.
J.} \textbf{71}, 2356 (1996); M. D. Wang \textit{et al.}, \textit{Biophys.
J.} \textbf{72}, 1335 (1997); A. D. Mehta, K. A. Pullen and A. Spudich,
\textit{FEBS Lett.} \textbf{430}, 23 (1998); A. D. Mehta, M. Rief
and J. A. Spudich, \textit{J. Biol. Chem.} \textbf{274}, 14517 (1999). 
\bibitem{Molecules_2}T. E. Fisher, P. E. Marszalek and J. M. Fernandez, \textit{Nat. Struct.
Biol.} \textbf{7}, 719 (2000); T. E. Fisher \textit{et al.}, \textit{Trends
Biochem. Sci.} \textbf{24}, 379 (1999); M. Carrion-Vazquez \textit{et
al.}, \textit{Prog. Biophys. Mol. Biol.} \textbf{74}, 63 (2000). 
\bibitem{tecnica2}L.N.Epele, H. Fanchiotti, C.A. Garcia Canal and M. Marucho, \textit{Phys.
Lett.} \textbf{B523}, 102 (2001). 
\bibitem{tecnica3}L.N.Epele, H. Fanchiotti, C.A. Garcia Canal and M. Marucho, \textit{Phys.
Lett.} \textbf{B556}, 87 (2003). 
\bibitem{tecnica1}L.N.Epele, H. Fanchiotti, C.A. Garcia Canal and M. Marucho, \textit{Nucl.
Phys.} \textbf{B583}, 454 (2000). 
\bibitem{abramovich}M. Abramowitz and I. Stegun, \emph{Handbook of Mathematical Functions}
(Dover, New York, 1970). 
\bibitem{hyper}H. Bateman, \emph{Higher Transcendental Functions} (Mc Graw Hill,
New York, 1953). 
\bibitem{Wang}M. C. Wang and E. Guth, \textit{J. Chem. Phys.} \textbf{20}, 1144
(1952).
\bibitem{Hs}C. Hsiung, H. Hsiung and A. Gordus, \textit{J. Chem. Phys.} \textbf{34},
535 (1961).
\bibitem{norman}Norman Bleistein and Richard Handelsman, \emph{Asymptotic expansion
of integrals} (Dover, New York, 1986). 
\bibitem{copson}E.T. Copson, \emph{Asymptotic Expansions} (Cambridge University Press,
1965). 
\bibitem{jefrey}H. Jeffrey and B.S. Jeffrey, \emph{Methods of Mathematical Physics}
(Cambridge University Press, , 1966). 
\bibitem{tablarusa}I.S. Gradshteyn and I.M. Ryzhik, \emph{Table of Integrals, Series,
and Products} (Academic Press, New York, 2000). 
\end{thebibliography}
\end{document}